\documentclass[journal]{IEEEtran}
\usepackage{cite}

\ifCLASSINFOpdf
\else
\fi
\usepackage{graphicx}
\usepackage{eurosym}
\usepackage{bbm}

\usepackage{amsmath}
\usepackage{amssymb}

\usepackage{tabu}

\begin{document}
%
\title{Measuring Player Retention and Monetization using the Mean Cumulative Function}
%
%
%

\author{Markus Viljanen*, Antti Airola, Anne-Maarit Majanoja, Jukka Heikkonen, Tapio Pahikkala
\thanks{The authors are with the Department of Future Technologies, University of Turku, 20014 Turku, corresponding author email: majuvi@utu.fi.}
}

%
%

{}
%



\maketitle

\begin{abstract}

Game analytics supports game development by providing direct quantitative feedback about player experience. Player retention and monetization in particular have become central business statistics in free-to-play game development. Many metrics have been used for this purpose. However, game developers often want to perform analytics in a timely manner before all users have churned from the game. This causes data censoring which makes many metrics biased. In this work, we introduce how the Mean Cumulative Function (MCF) can be used to generalize many academic metrics to censored data. The MCF allows us to estimate the expected value of a metric over time, which for example may be the number of game sessions, number of purchases, total playtime and lifetime value. Furthermore, the popular retention rate metric is the derivative of this estimate applied to the expected number of distinct days played. Statistical tools based on the MCF allow game developers to determine whether a given change improves a game, or whether a game is yet good enough for public release. The advantages of this approach are demonstrated on a real in-development free-to-play mobile game, the Hipster Sheep.

\end{abstract}


%
\IEEEpeerreviewmaketitle

\section{Introduction}
%
%
%
%
Digital industry is a major sector of the modern economy. Products and services sold digitally are a source of revenue for many companies, the game industry in particular has grown to form a large part of the app economy. The freemium business model, the proliferation of mobile devices and the expanding possibilities in data gathering have all shaped the recent evolution. These developments have made player retention alongside the resulting monetization a central development target and analytics benchmark. Indeed, many companies have come to regard them as core metrics of profitability and have implemented tracking as a part of their everyday business activities \cite{Seufert2014}. Because there are significant adverse consequences to making decisions based on inadequate metrics, research is required to investigate if the current game metrics are as efficient and informative as needed.

To address the challenge of obtaining insight from player data, a set of free-to-play metrics has been proposed \cite{Fields2013}. Metrics are well-defined calculations that aggregate the game data to a single statistic. 
The total number of sessions, distinct days played, purchases, total playtime, total revenue, etc. per player are informative measures of total time and money spent. As an estimate of the expected population value, the mean value of these is an important aggregate metric. The expected Lifetime Value (LTV) could be argued to be the most important business statistic, because the return on investment is given as the difference of the LTV and the acquisition cost of a player \cite{Sifa2015purchase}. 
Likewise, the expected playtime can be used to measure how much players enjoy the game \cite{Sifa2014}.

Game analytics that is provided to game developers often places high time-demands on game metrics. The game developers for example can rarely afford to wait to the point when every player has churned, i.e. quit playing the game, in order to measure the total playtime. In addition, players come and go over time which results in different observation times for each player. This phenomena is often referred to as data censoring. The presence of censoring has resulted in industry metrics that emphasize computability as the data comes in. The retention rate in particular has become a central industry metric of player engagement \cite{Fields2013}. Given a day that a group of players started playing the game, it computes the percentage of these players that return to the game in every subsequent day. It is common to use retention rate on a given day (say, first day, week or month) as a benchmark. 

In this work, we show that many academic game metrics can be estimated even for censored data. The proposed approach is based on accepted methods from reliability engineering and biostatistics, which deal with similar kind of data: recurrent events and associated costs. A tool known as the mean cumulative function can be used to estimate the expected number of sessions, number of purchases, total playtime and lifetime value over time. In particular, when it is applied to the expected number of distinct days played, the derivative in fact corresponds to the retention rate metric. It generalizes the ordinary retention rate in one sense, since one can use players with different censoring lengths and extends it to a continuous time domain by using the number of sessions. The MCF therefore offers a novel way to approach the problem of measuring, visualizing and analyzing player retention and monetization in games. 
When we apply the method, we are also able to answer the following scientific questions: 
\begin{itemize}
\item What is the expected value at time $T$?
\item What is the uncertainty of the estimate?
\item Which is better of the versions A and B?
\item What is the confidence that the version is better?
\end{itemize}


The paper is structured as follows. Section II reviews background literature introducing the current metrics and the MCF. Section III introduces the research objectives and research data. Section IV describes the MCF as a game analytics method and Section V analyzes the advantages of using the MCF for gaming data. Section VI concludes our findings.

\section{Literature Review}

\subsection{Game Analytics}
Academic literature understands ``retention'' as an umbrella term that together with churn has been used to study many engagement metrics \cite{Feng2007,Tarng2008,Chen2009,Weber2011using,Debeauvais2014,Debeauvais2015,Sifa2014,viljanen2017,Allart2016,Isaksen2015,Saas2016,viljanen2016modeling,viljanen2017abc,Hadiji2014,Rothenbuehler2015,Runge2014,Tamassia2016}. Studies have for example treated retention as a metric that measures various aspects of player activity \cite{Feng2007,Tarng2008}, session time \cite{Chen2009}, total sessions \cite{Weber2011using}, total purchases \cite{Debeauvais2015}, playtime \cite{Sifa2014,viljanen2017,Allart2016}, gates cleared \cite{Debeauvais2015,Isaksen2015}, days active \cite{Saas2016}. In contrast, the term retention often refers to the specific ``retention rate'' metric in the industry \cite{Seufert2014}. A case study elaborating this and other industry metrics has been published \cite{Fields2013}, and it has also been studied in the academia \cite{viljanen2016modeling, viljanen2017abc}.

Gaming literature has not simply measured these metrics, but also sought to develop predictive models and interpret the effect of player features on them. Linear \cite{Weber2011using,Debeauvais2015} and the Cox \cite{Allart2016,Saas2016,viljanen2017abc} regression have been applied to study the effect of covariates to retention metrics. Churn is a natural complement of retention, since users churning at time $T$, regardless of the definition of time, implies users were retained for time $T$. Churn prediction has been an especially active research area on predictive models. Many machine learning models have been used and contrasted to the simpler tools of Logistic regression \cite{Chen2009,Debeauvais2014,Debeauvais2015,Hadiji2014,Rothenbuehler2015,Runge2014,Tamassia2016} and especially Hidden Markov Models, \cite{Hadiji2014,Rothenbuehler2015,Runge2014,Tamassia2016} for player churn classification.

The research on metrics, regression models and involved predictive machine learning models is largely complementary. A game analyst’s first concern is often to understand players and game performance as reported in analytics dashboards through simple and appropriate metrics. The next stop might be to understand how general principles, marketing and game design affect player retention. When a well-defined predictive task can be formulated that impacts revenues and requires as accurate answers as possible, for example the recommendation of games to players, a natural solution is then to implement a sophisticated machine learning model. The contribution of this study belongs to the research on metrics in game analytics.

\subsection{Reliability Engineering and Biostatistics}

The problem of measuring and understanding player type data is shared by reliability engineering and medical literature. Therein one deals with machine failures or patient syndromes, possibly with associated costs, as recurrent events. Reliability has traditionally studied parametric, often single process models \cite{lawless1983} with medicine emphasizing nonparametric methods for several subjects \cite{andersen85}. Similar to how metrics and interpretative models can be contrasted to predictive models in games, the emphasis between the two fields often varies. Parametric models can be used to predict outside the data set, which is important for example in maintenance planning, whereas medical study often uses nonparametric methods to understand how drugs and other interventions impact patient survival. Another main reason for this divergence is in the nature of the data sets; simple laws often describe the homogeneous and independent nature of failing machine parts, whereas human subjects undergoing a life cycle often vary both as individuals and display time-inhomogeneity. In this regard gaming seems to be closer to medicine. Nevertheless, parametric models have been used in marketing for purchase processes \cite{Fader2009}, of which the BG/NBD model has been found to struggle with free-to-play games \cite{Hanner2015Counting}.

\subsection{Robust Methods for Recurrent Events}

The Mean Cumulative Function (MCF) \cite{nelson1995} estimate of the population mean is a central concept in both reliability and biostatistics, where it forms the foundation of more advanced statistical methods. Subsequent to its introduction, research effort in these fields has been applied to the robust analysis of recurrent events as described in the review of Lawless \cite{lawless1995} and the book by Cook and Lawless \cite{cook2007}. This ``robustness'' makes these methods applicable to gaming data.

Robustness in this context means that the method is free of the Poisson process assumption that is implicit in a large part of recurrent event and cumulative cost literature. This assumption is often employed for simplicity and because it allows many theoretical results to be derived \cite{Kingman2005}. The assumption says that the player behavior is independent of the player history. However, because of player churn it is easy to deduce that the Poisson assumption cannot apply as follows.

The decreasing retention rate in a given player cohort, for example, has two major causes: players quit playing and they become less excited of the game as time goes by. If only the latter effect occurred and there was negligible variation between the players, assuming that players play according to a Poisson process might be a good approximation. However, because players churn over time the play rate is not sufficient to describe the players. For example, a play rate of 0.2 sessions per day in a cohort does not mean that every player has the independent play intensity of 0.2 sessions/day, but that there are some players who have quit with a constant intensity of 0 sessions/day, and some players with an actual intensity of playing, say 1.0 sessions/day in the 20\% still active. Said another way, since past player inactivity predicts future player inactivity because they are more likely to have churned, player behavior is not independent of history. In this paper we therefore drop this assumption and use robust methods.

\subsection{Research on The Mean Cumulative Function}

We finally briefly review the literature in which the theory is developed to provide references to the formulas and extensions mentioned in the paper. 

The MCF was originally introduced by Nelson \cite{nelson1969,nelson1972Theory} and Altschuler \cite{Altshuler1970Theory} independently as a way to study the cumulative failure intensity in a single event process and subsequently studied in depth through a recurrent event framework based on counting processes by Aalen \cite{aalen1978nonparametric}. Nelson also noted \cite{nelson1988Graphical,nelson1995} that the estimator can be used in general settings to estimate cumulative cost. Cook and Lawless \cite{cook2007} discuss additional results and applications of cumulative cost.

The uncertainty in the MCF estimate may be derived with or without the Poisson assumption. 
Confidence intervals are pointwise estimates, which means that they quantify the probable range of the MCF at a given time point $t$. Confidence bands are required to estimate the area which is likely to enclose the MCF at every time point $t$. Aalen \cite{aalen1978nonparametric} discussed confidence intervals for a counting process based on the Poisson assumption, and Poisson confidence bands were later discussed in both reliability \cite{Vallarino1988confidence} and biostatistics \cite{Bie1987confidence}. The robust MCF confidence intervals were introduced in the applied research of Nelson and Doganaksoy \cite{nelson1989program}, Robinson \cite{robinson1990standard,robinson1995standard}. They differ slightly in that Nelson uses unbiased and Robinson positive but biased variance estimates. This small difference is analogous to how the sample variance can be estimated with either $n$ or $n-1$ in the denominator, known as the Bessel's correction \cite{upton2014dictionary}. These intervals, among other results, are discussed in Lawless and Nadeau \cite{lawless1995nonparametric}. Robust confidence bands are introduced in Lin et al. \cite{lin2000semiparametric}.

Doganaksoy and Nelson \cite{doganaksoy1991method,doganaksoy1998method} presented a simple test for comparing two MCFs pointwise. This test is based on the difference between the two estimates, which is an unbiased estimate of the actual difference. To compare multiple cohorts, a pairwise test between all MCFs can be performed but the confidence intervals should then be corrected for a k-sample comparison, i.e. the chance of the event that some of the comparisons differ significantly. Nelson \cite{nelson2003recurrent} elaborated how this is achieved using the Bonferroni correction, a simple conservative inequality, or analysis-of-variance results.

To perform a comparison of equality between MCFs, instead of a simple test of pointwise differences, many nonparametric linear rank tests have been presented in the biostatistics literature to compare two- or k-sample data. Andersen et al. \cite{andersen1982linear} showed that the comparisons could be combined in the counting process framework as k-sample tests using weight functions. The robust two-sample test was introduced by Pepe and Cai \cite{pepe1993some}, Lawless and Nadeau \cite{lawless1995nonparametric}. Cook et al. \cite{cook1996robust} presented a $k$-sample comparison and compared weight functions. These results were later discussed rigorously in the empirical process framework of Lin et al. \cite{lin2000semiparametric}.

The famous Cox regression model \cite{andersen1982cox} establishes semiparametric regression in this setting. Cook \cite{cook2007} reviews the robust procedures for this model. The model could be used to extend the simple analysis provided by the MCF, it for example has been used in gaming as a recurrent event ABC-test \cite{viljanen2017abc}.

\begin{figure}[h]
\centering
\includegraphics[width=0.8\linewidth]{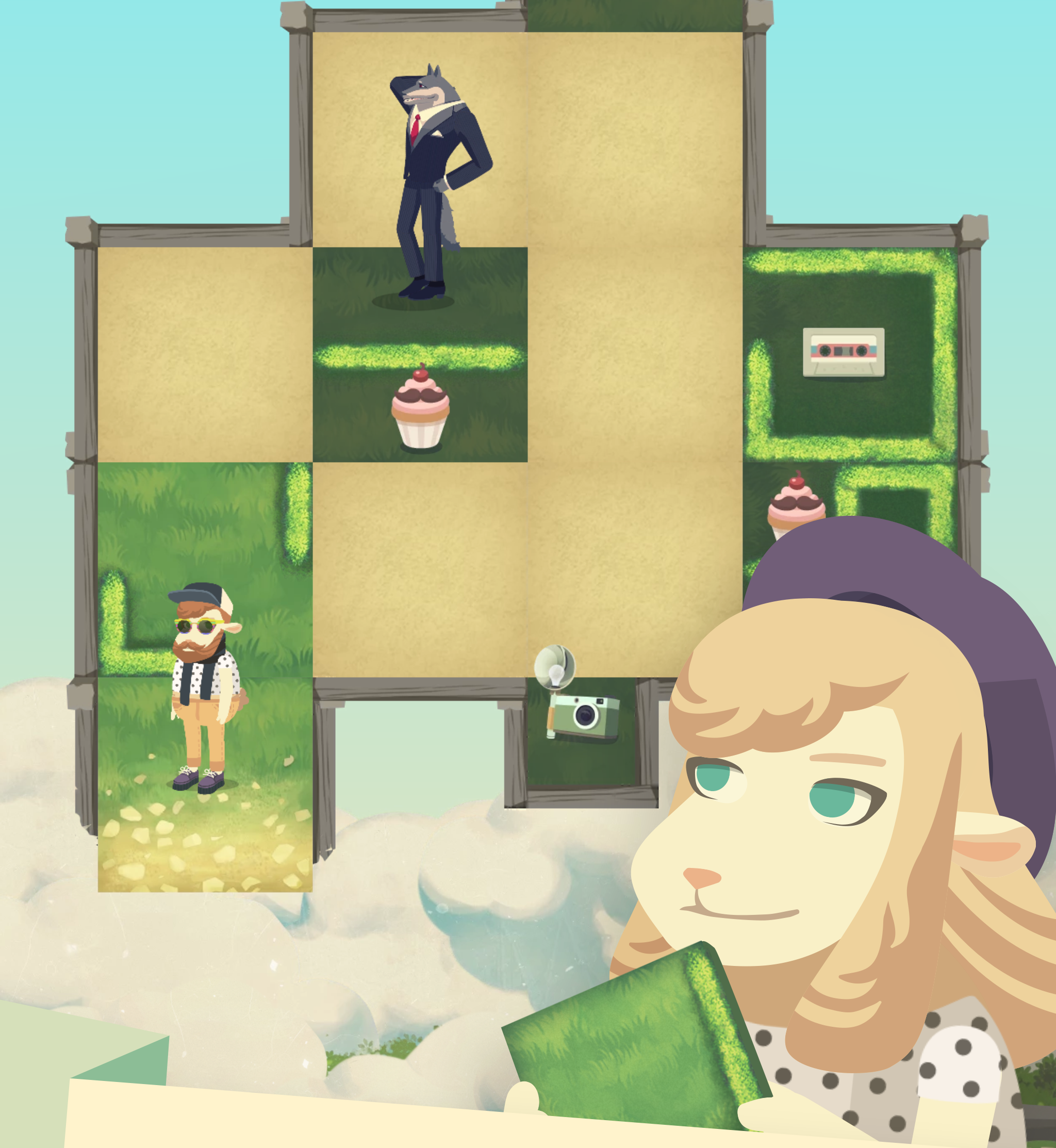}
\caption{Hipster Sheep screenshot used with the permission of Tribeflame Ltd.}
\label{fig:hipster_screenshot}
\end{figure}

\newcommand{\dset}{D}
\newcommand{\evt}{t}
\newcommand{\cind}{y}
\newcommand{\rvcost}{C}
\newcommand{\cost}{c}
\newcommand{\dcost}{\Delta c}
\newcommand{\cevnum}{N}
\newcommand{\devnum}{dN}

\begin{table}
\centering
\caption{Data set example: Hipster Sheep players}
\begin{tabu} to \linewidth {|X[0.1]|X[0.1]|X[0.8]|X[0.3]|X[0.3]|X[0.2]|}
\hline
ID & N & Timestamp & Time (d) & Type & Value \\
\hline
1 & 1 & 2016-07-29  01:34:33 & 0.00 & session & 08:45 \\
1 & 2 & 2016-07-29  03:38:59 & 0.09 & session & 12:52 \\
1 & 3 & 2016-07-29  03:51:25 & 0.10 & purchase & 1.09\euro \\
1 & 4 & 2016-07-30  04:26:04 & 1.12 & session & 29:10 \\
1 & 5 & 2016-07-30  15:32:13 & 1.58 & session & 00:01 \\
1 & 6 & 2016-08-14  14:18:30 & 16.53 & session & 15:32 \\
1 & 7 & 2016-11-09  00:00:00 & 102.93 & censored & \\
2 & 1 & 2016-09-08  13:20:17 & 0.00 & session & 04:37 \\
2 & 2 & 2016-09-08  14:07:40 & 0.03 & session & 04:40 \\
2 & 3 & 2016-09-08  14:24:31 & 0.04 & session & 00:01 \\
2 & 4 & 2016-09-10  14:05:10 & 2.03 & session & 03:17 \\
2 & 5 & 2016-10-04  19:48:40 & 26.27 & session & 00:12 \\
2 & 6 & 2016-10-04  19:48:55 & 26.27 & session & 02:34 \\
2 & 7 & 2016-10-06  13:17:42 & 28.00 & session & 00:03 \\
... & ... & ... & ... & ... & ...\\
\hline
\end{tabu}
\label{table:hipster_players}
\end{table}

\section{Research Methods and Data Set}

The research approach included the following phases.

Literature review: based on research in reliability engineering and biostatistics, we collected the robust methods required for game analytics in the previous chapter with the goal of introducing the MCF as a new tool for game analytics.

Method development and implementation: we first review the current approach to measuring retention and monetization under censoring and then introduce the MCF as a new approach. We demonstrate this development with an actual data set. We show that the MCF is closely related to the rate type metrics in the industry, and it allows the expected value of academic game metrics to be estimated under censoring.

Comparison to existing methods: the final section argues, using several real world game development problems we have encountered, that the MCF has unique advantages over existing metrics in interpretation of game quality and player behavior.

The data set we use is from a free-to-play mobile game called Hipster Sheep being developed by Tribeflame Ltd. The game is a casual puzzle game where the player guides a human-like sheep through interactive labyrinths with collectibles, cameras and wolves.  Like many modern free-to-play games, the game blends skill and luck in order to entice the player to in-game purchases. The game is targeted at young adult females and has a lighthearted artistic theme reflecting the Hipster lifestyle. The picture of Figure~\ref{fig:hipster_screenshot} probably tells more than a thousand words. 

Table~\ref{table:hipster_players} shows an excerpt of this data. In the example, two players (ID) both have 7 events (N) that are recorded in calendar time (Timestamp) and time since install (Time) in days. The events belong to one of three categories (Type): sessions, purchases and censoring indicator. Censored means that player history from this point forward is unknown. The sessions have associated lengths and the purchases have associated purchase amounts, which are shown as the event values (Value).

\begin{figure}[t]
\centering
\includegraphics[width=1.0\linewidth]{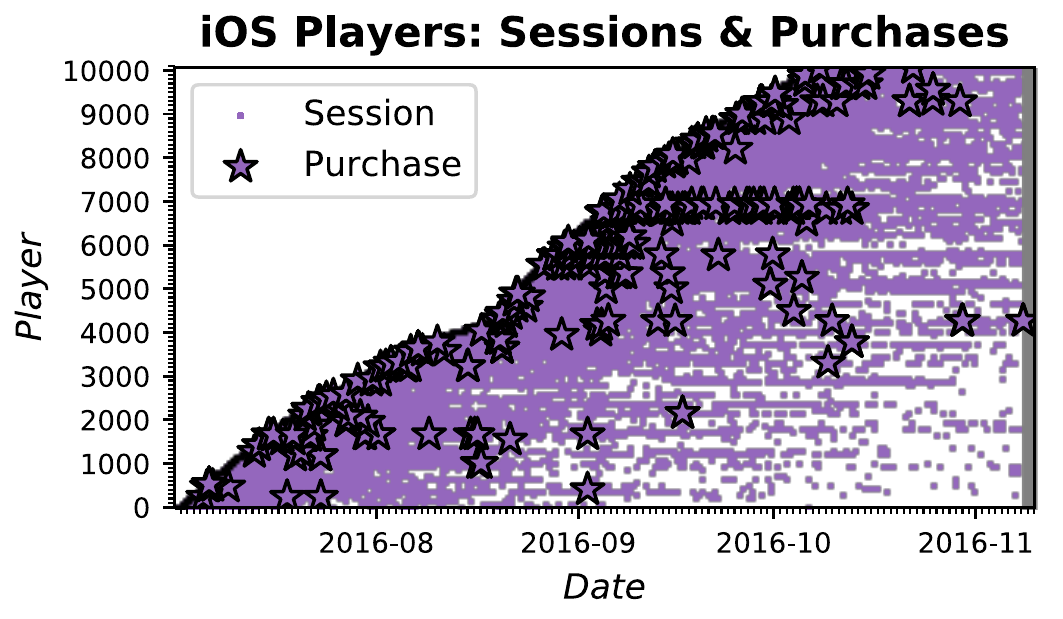}
\caption{Timelines of individual players in Hipster Sheep development: players are plotted on y-axis with resulting purchases and recurrent sessions on x-axis.}
\label{fig:players_population}
\end{figure}

\section{Research Results}

\subsection{Current Approach: Using the Retention Rate}

The need for metrics is vividly illustrated in Figure~\ref{fig:players_population}, where a scatter plot shows sessions and purchases of 10 000 iOS players acquired for a Beta test in Hipster Sheep. If the data was not censored by the maximum observable day in the data set, popular academic metrics like playtime per player could be obtained by summing over time on x-axis. Other natural metrics obtained in such a way would be the number of sessions or purchases, and the total revenue per player. If the data consists of incomplete user life cycles as is the case here, summing results in downward biased estimates because not all players have churned and their playtime is larger than what we have observed so far. Alternatively, many popular free-to-play retention and monetization metrics can be obtained by summing over the the players on y-axis using some binning frequency on x-axis \cite{Seufert2014}, resulting in the following popular industry metrics, for example:
\begin{itemize}
\item Daily New Users (DNU): new users per day.
\item Daily Active Users (DAU): active users per day.
\item Retention Rate (RR): active users per day, relative to the first day the users played at.
\item Average Revenue per User (ARPU): revenue per user in a given time period such as one day.
\end{itemize}

We have plotted new users, active users and purchases in Figure 3, where we see that the underlying retention rate of the new users clearly determines the resulting user activity. Because the retention rate of new users tells us how players enjoy the game over time and it explains an important factor of game success, industry uses this statistic to understand how the game is performing \cite{Seufert2014}. The DAU is a cruder metric in terms of user engagement \cite{Fields2013,Seufert2014,Fader2009} because it ignores the temporal pattern of how the game retains the players by having both new and old players which convolutes the trend we are trying to understand. Fader and Hardie \cite{Fader2009} for example are careful to distinguish aggregate and cohort metrics. Aggregate metrics are obtained as the current population status, whereas cohort metrics like the retention rate describe the player behavior. Cohort metrics like the play or purchase frequency relative to the install date are more informative in this regard \cite{Seufert2014,Fader2009}.

\begin{figure}[t]
\centering
\includegraphics[width=1.0\linewidth]{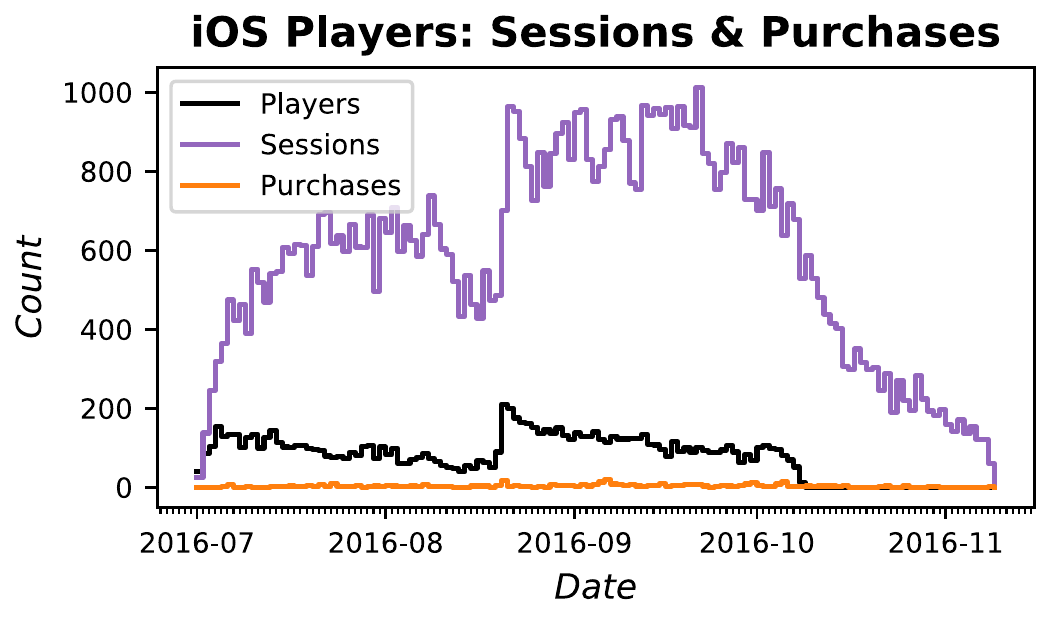}
\caption{Total sessions and purchases of players in Figure 1. are the sum over all acquisition day based cohort session and purchase rates, multiplied by the number of new players in each cohort. The cohort retention pattern is unclear.}
\label{fig:stats_population}
\end{figure}

To illustrate this approach, Table~\ref{table:hipster_retention} counts the total sessions for some of these player cohorts. The retention rates are calculated using the rows in the table, which group the players according to each starting day into cohorts. For each cohort, the number of sessions per day is computed for the days following the starting day. The rate of sessions per day is the number of sessions divided by the total number of players. For example, in the ``3.7'' row we had 103 new users who played 176 sessions during the first day. The one day (1d) retention rate is 176/103=1.7 sessions per day. We further obtain the 2d to 5d rates as a decreasing sequence: 1.0, 0.5, 0.4, 0.2 ... This calculation is also commonly done with the total number of sessions replaced by the total number of unique users. One then obtains the percentage of users that return to the game. 


\begin{table}
\caption{Retention example: recurrent sessions by acquisition day}
\begin{tabu} to \linewidth {|X|X|X|X|X|X|X|X|X|}
\hline
\multicolumn{2}{|c|}{Cohort} & \multicolumn{7}{|c|}{Play Day} \\
\hline
Start Day & DNU & 1.7 & 2.7 & 3.7 & 4.7 & 5.7 & 6.7 & 7.7 \\
\hline
7.7 & 134 & & & & & & & 152 \\
6.7 & 134 & & & & & & 194 & 86 \\
5.7 & 128 & & & & & 159 & 146 & 106 \\
4.7 & 153 & & & & 166 & 121 & 65 & 48 \\
3.7 & 103 & & & 176 & 107 & 54 & 44 & 20 \\
2.7 & 85 & & 105 & 56 & 38 & 25 & 21 & 8 \\
1.7 & 39 & 24 & 32 & 13 & 8 & 5 & 4 & 1 \\
\hline
DAU & & 24 & 137 & 245 & 319 & 364 & 474 & 421 \\
\hline
\end{tabu}
\label{table:hipster_retention}
\end{table}

We see that metrics can be aggregated over time to obtain a value per player or aggregated over players to obtain a value per time. Both approaches have their strengths and weaknesses. The player total based metrics do not explain how the total was accumulated over time, that is how the players were retained. To deal with censoring, we could simply compute the interim distribution for players that have the same censoring time, but we would then need to limit all players to this censoring time or exclude them. The rate based metrics on the other hand reveal the player retention pattern and the computed values do not change as the censoring window expands with new data, but point estimates such as 30d, 60d and 90d retention rates may leave out potential information present in other parts of the retention curve. 

In this paper, we demonstrate an elegant way to combine the metrics using the Mean Cumulative Function (MCF). We obtain the considerable benefit that the metrics then naturally generalize to data with varying censoring lengths, and statistical tests can also be performed with the full data set.

\subsection{New Approach: The MCF Estimate}

We now motivate the MCF as a model-free estimate of the population mean and recollect confidence intervals for this estimate. We then introduce two types of comparisons used in the medical literature for AB-tests: pointwise and complete. The mathematical results are accompanied by illustrations using the data set of Hipster Sheep. We define the estimate following Cook and Lawless \cite{cook2007} for a concise formula that extends to several types of time domains. Nelson \cite{nelson2003recurrent} provides an introduction with an applied focus.

\begin{figure}[t]
\centering
\includegraphics[width=1.0\linewidth]{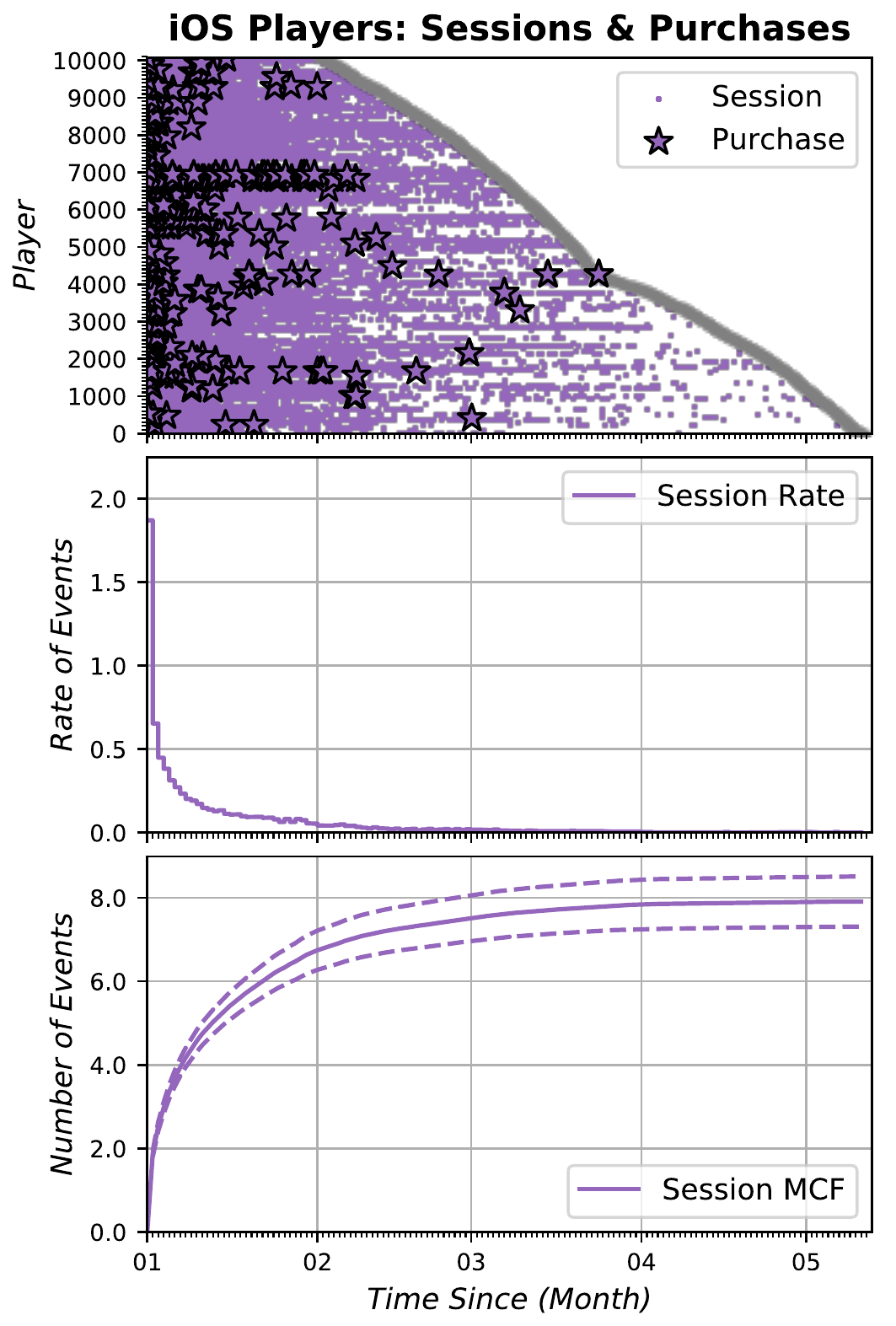}
\caption{Players are plotted by time since the first session in the topmost figure. On the same axis the middle figure provides a piecewise session rate estimate and the last figure is the MCF with 95\% confidence intervals.}
\label{fig:mcf_population}
\end{figure}

Figure~\ref{fig:mcf_population} plots sessions and purchases as a time shifted version of Figure~\ref{fig:players_population}, where we use time since install to measure player retention. The resulting observation limits are denoted by the gray diagonal line, where the players are observable between their first session and the data collection limit.

\newcommand{\pc}{m}
\newcommand{\tc}{n}

Given such a set of observed data, we want to estimate the amount of money spent or time played by a player over time. In statistical terms, the population is described by a random variable $\rvcost(\evt)$ denoting the cumulative cost for a player up to time $t$ since starting the game. We
calculate an estimate $\widehat{\mathbb{E}}[\rvcost(\evt)]$ of the expected cost $\mathbb{E}[\rvcost(t)]$ from a sample of censored player event histories. These consist of times and costs for events such as sessions or purchases, as well as the censoring times.
For example, let us consider the player histories in Table~\ref{table:hipster_players}. Player ID 1 has five session events, happening at times 0.00, 0.09, 1.12, 1.58, 16.53 in days. The censoring time 102.93 denotes the length of time from the start of first session to when the data was gathered. The cost of each event is one when estimating the cumulative number of sessions, and equal to session length when estimating cumulative playtime. Event times, costs and censoring times are analogously defined for the other players.

Formally, let $\dcost_i(\evt_j)$ denote the cost of an event happening to player $i$ at time point $\evt_j$, where $1\leq i\leq\pc$ and $1\leq j\leq\tc$ are the indices of players and distinct event times, respectively.  Further, each player has a censoring time $\tau_i$ after which the events are not known. Let $\cind_i(\evt)$ denote whether the player is observable at time $\evt$, meaning that $\cind_i (\evt)=[\evt \leq \tau_i]$. The $\dcost_i(\evt_{j})$ is defined as zero when there is no event happening to player $i$ at time $\evt_j$, or when the player is censored, i.e. $\cind_i(\evt_j)=0$. Finally, we denote by $\cost_i(\evt) = \sum_{j:\evt_{j} \leq \evt} \dcost_i(\evt_{j})$ the total cost accumulated by player $i$ in the interval $(0,\evt]$.

To define the aggregate values over all players, denote by $\dcost_{\circ}(\evt) = \sum_{i=1}^{m} \dcost_i (\evt)$ the total cost and by $\cind_{\circ}(\evt) = \sum_{i=1}^{m} \cind_i(\evt)$ the total number of players observable at time $t$. The Mean Cumulative Function (MCF) estimates the expected value $\mathbb{E}[\rvcost(\evt)]$ as cumulative sum of total costs over all event times, taking into account the number of players observable: 
\begin{align}\label{robest}
\widehat{\mathbb{E}}[\rvcost(\evt)]=\sum_{j:\evt_j\leq \evt}\frac{\dcost_{\circ}(\evt_j)}{\cind_{\circ}(\evt_j)}
\end{align}

In the special case where the total cost $\rvcost(\evt)$ is the number of events that have occurred, the MCF also known as the Nelson-Aalen estimate. The MCF estimate of the expected number of sessions in the data set of Figure~\ref{fig:players_population} is illustrated in Figure~\ref{fig:mcf_population}. 

The rate of events is given by the instantaneous change in the expected number of events, or the derivative $\frac{\partial}{\partial \evt} \widehat{\mathbb{E}}[\rvcost(\evt)]$. Smooth approximations of the rate can therefore be estimated from data (see e.g. \cite{cook2007} for the details). A simple approximation that assumes a constant derivative per day is plotted in the center of Figure~\ref{fig:mcf_population}. This derivative is in fact a generalization of the continuous retention rate in Table~\ref{table:hipster_retention} to censored data. If instead we estimate the expected number of distinct days played, the estimate $\widehat{\mathbb{E}}[\rvcost(\evt)]$ increments on discrete times $t=1,2,...$ and the finite differences $\frac{\partial}{\partial \evt} \widehat{\mathbb{E}}[\rvcost(\evt)]=\widehat{\mathbb{E}}[\rvcost(\evt)] - \widehat{\mathbb{E}}[\rvcost(\evt-1)]$ correspond to the ordinary retention rate. The estimate based on purchases and associated profits is the expected lifetime value per player and the derivative is the instantaneous revenue per player over time. The MCF estimate therefore combines the cohort rate based metrics as the derivatives of the corresponding expected mean value, which can be estimated using a robust model-free method up to the limit we have data for. 

We assumed that the data set is a sample of independent and identically distributed realizations of the random variable $\rvcost(\evt)$. The dataset itself is therefore a random variable and so is the estimator based on it. For example, from millions of potential players we have many ways of choosing a sample of 10 000, and by chance we may obtain players that behave differently from most players in the population. The metrics we compute based on the sample could therefore be different from the metrics we are trying to measure. Any estimate based on a sample therefore has sampling uncertainty associated with it. This uncertainty can be quantified with confidence intervals.

To compute the 95\% confidence intervals displayed in Figure~\ref{fig:mcf_population}, we proceed as follows. Note that at every time $\evt$, the term $\widehat{\mathbb{E}}[\rvcost(\evt)]$ is a sum of independent random variables, i.e. the player values, and by the central limit theorem it therefore is asymptotically normally distributed. Using a normal distribution, the $95\%$ confidence intervals are
$\widehat{\mathbb{E}}[\rvcost(\evt)] \pm z\sqrt{\widehat{\textnormal{Var}}[\widehat{\mathbb{E}}[\rvcost(\evt)]]}$ with $z=1.96$. A robust variance estimate required by the formula can be shown to equal \cite{cook2007}:
\begin{align}\label{robvarest}
\widehat{\textnormal{Var}}[\widehat{\mathbb{E}}[\rvcost(\evt)]]
=\sum_{i=1}^\pc\left(\sum_{j:\evt_j\leq \evt}\frac{\cind_i(\evt_j)}{\cind_{\circ}(\evt_j)}\left(\dcost_i(\evt_j)-\frac{\dcost_{\circ}(\evt_j)}{\cind_{\circ}(\evt_j)}\right)\right)^2
\end{align}

Without censoring these estimates are simpler because it becomes possible to use standard techniques. For example, an unbiased estimate of the expected value at time $\evt$ is then the sample mean $\widehat{\mathbb{E}} [\rvcost(\evt)] = \sum \cost_i (\evt)/\pc$ and the variance may be estimated by the (biased) sample variance $\widehat{\textnormal{Var}}[\rvcost(\evt)] = \sum(\cost_i(\evt) - \widehat{\mathbb{E}}[\rvcost(\evt)])^2 / \pc$. These formulas can also be derived from the formulas (\ref{robest}) and (\ref{robvarest}) by substituting a common censoring time for all players. The estimates therefore have the nice property that they equal the well-known estimators in the special case of uncensored data.

\begin{table}[b]
\caption{MCF example: the number of sessions}
\begin{tabu} to \linewidth {|X|X|X[0.7]|X|X|X|X|}
\hline
Time $\evt_j$
&Observ. $\cind_{\circ}(\evt_j)$
&Cost $\dcost_{\circ}(\evt_j)$
&Incr. $\frac{\dcost_{\circ}(\evt_j)}{\cind_{\circ}(\evt_j)}$
&MCF $\widehat{\mathbb{E}}[\rvcost(\evt_j)]$
&MCF LCI
&MCF UCI\\
\hline
0,00000&0&0&0,00000&0,00000&0,00000&0,00000\\
0,00003&10068&8&0,00079&0,00079&0,00024&0,00135\\
0,00005&10068&4&0,00040&0,00119&0,00052&0,00187\\
0,00006&10068&2&0,00020&0,00139&0,00066&0,00212\\
0,00007&10068&3&0,00030&0,00169&0,00089&0,00249\\
0,00009&10068&4&0,00040&0,00209&0,00119&0,00298\\
0,00010&10068&3&0,00030&0,00238&0,00143&0,00334\\
0,00012&10068&5&0,00050&0,00288&0,00183&0,00393\\
0,00013&10068&6&0,00060&0,00348&0,00233&0,00463\\
0,00014&10068&8&0,00079&0,00427&0,00300&0,00554\\
...&...&...&...&...&...&...\\
\hline
\end{tabu}
\label{table:hipster_mcf}
\end{table}

Finally, we illustrate the preceding discussion with a simple example of how to compute the MCF estimate of Figure~\ref{fig:mcf_population} using Table~\ref{table:hipster_mcf}. At every time point $\evt_j$ we denote by $\cind_{\circ}(\evt_j)$ the number of players observable and by $\dcost_{\circ}(\evt_j)$ the total number of sessions.
We have computed the increment and the accumulated sum, the MCF $\widehat{\mathbb{E}}[\rvcost(\evt)]$ estimate, at every row. Note that this sum could equally well be based on real-valued costs such as the session length or the purchase amount instead of the binary session increment at time $\evt_j$.

\begin{figure}[t]
\centering
\includegraphics[width=1.0\linewidth]{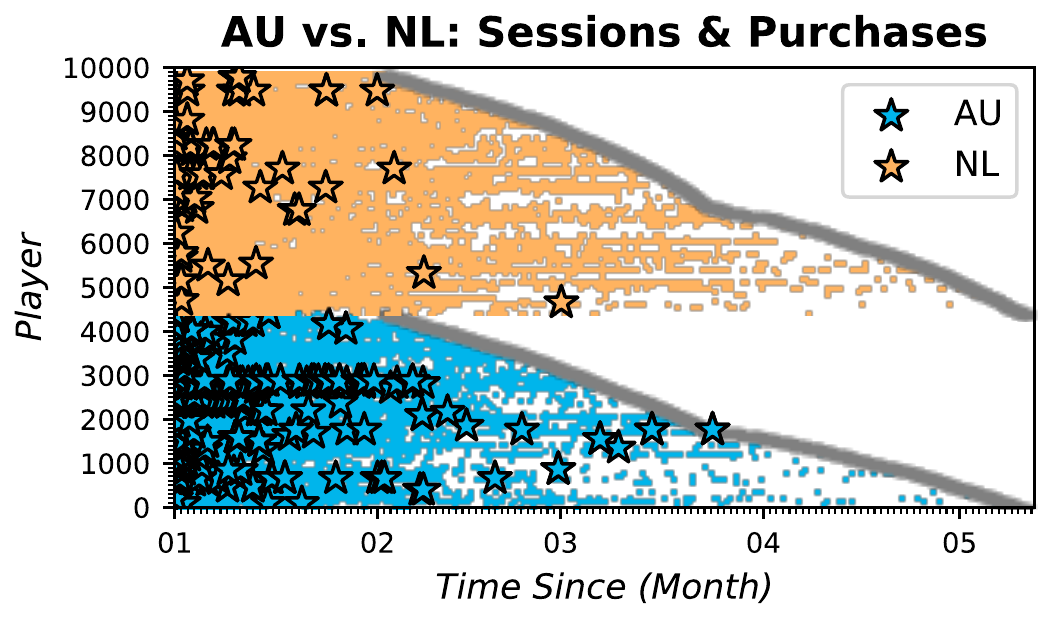}
\caption{Player timelines of two cohorts based on the player country of origin: Australia (AU) and Netherlands (NL) are shown in different colors.}
\label{fig:mcf_comparison}
\end{figure}

\subsection{MCF Comparison}

Suppose for simplicity that the players can be divided into two cohorts based on a player feature. The user acquisition in Figure~\ref{fig:mcf_population} for example consisted of players in Australia (AU) and Netherlands (NL), except for a few hundred random players. Figure~\ref{fig:mcf_comparison} displays these players divided into two cohorts based on the country of origin. It would be of interest to be able to compare these cohorts for retention and monetization, for example to know which country to acquire users from in the future. These tests are called AB-tests in gaming literature \cite{Seufert2014}, and are also used in medicine to assess treatment effectiveness.

\begin{figure}[t]
\centering
\includegraphics[width=1.0\linewidth]{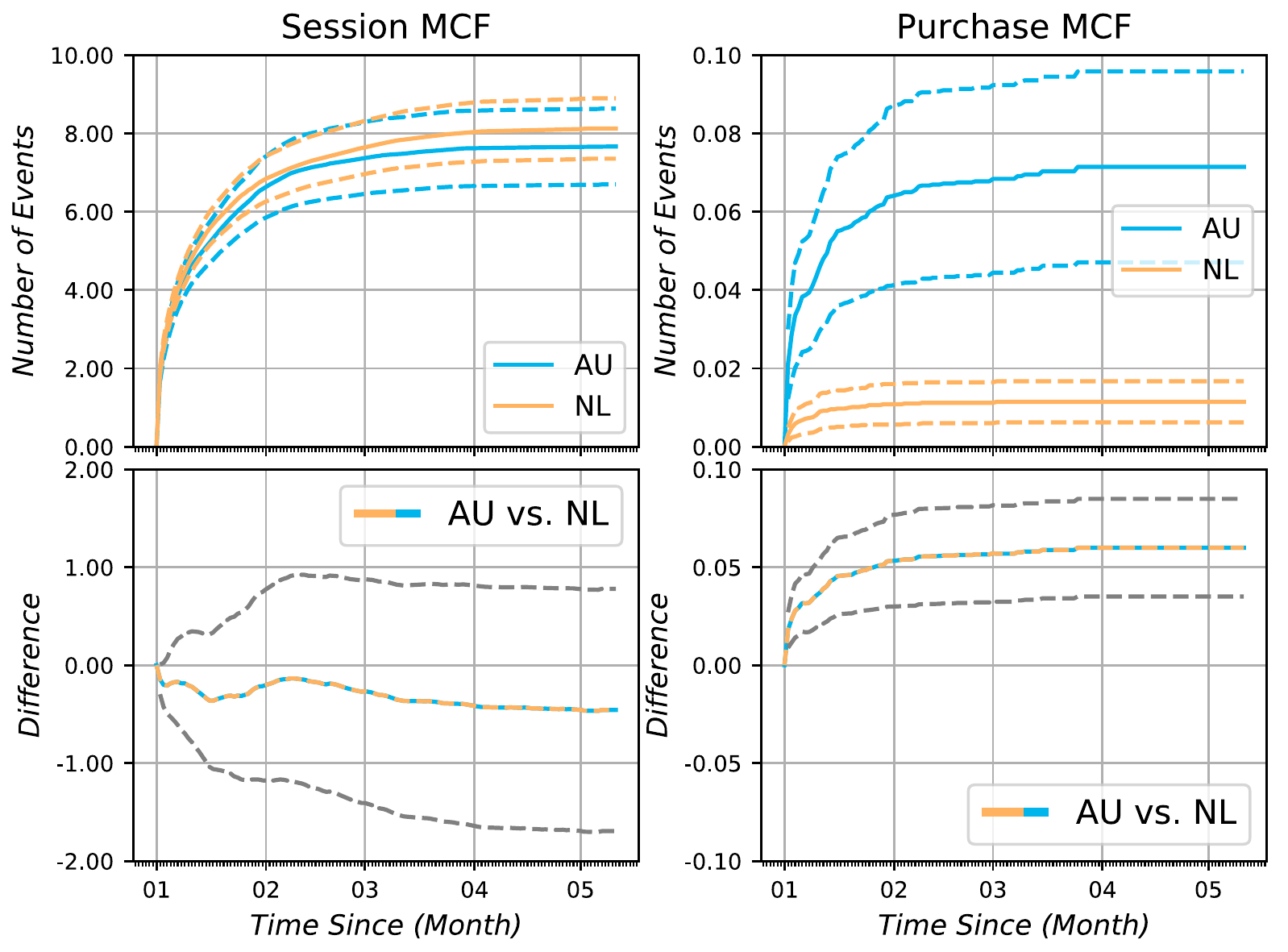}
\caption{The difference between players in Australia and Netherlands is not significant for session counts but is consistently significant for purchase counts using the pointwise comparison of MCFs since the intervals do not enclose zero.}
\label{fig:2x2}
\end{figure}

Denote the two cohort MCFs $\widehat{\mathbb{E}}[\rvcost_1(\evt)]$ and $\widehat{\mathbb{E}}[\rvcost_2(\evt)]$.
Since the sampling of players is assumed to be independent, the variance of the difference can simply be estimated as the sum of the MCF variances estimated using the previous formula. This results in the pointwise MCF difference estimate \cite{nelson2003recurrent}:
\begin{align}
\widehat{\mathbb{E}}[\Delta\rvcost(\evt)]=\widehat{\mathbb{E}}[\rvcost_1(\evt)]-\widehat{\mathbb{E}}[\rvcost_2(\evt)]
\end{align}
and the variance estimate of the pointwise MCF difference:
\begin{align}
\widehat{\textnormal{Var}}[\widehat{\mathbb{E}}[\Delta\rvcost(\evt)]]=\widehat{\textnormal{Var}}[\widehat{\mathbb{E}}[\rvcost_1(\evt)]]+\widehat{\textnormal{Var}}[\widehat{\mathbb{E}}[\rvcost_2(\evt)]]
\end{align}
The estimate of the difference $\widehat{\mathbb{E}}[\Delta\rvcost(\evt)]$ between player MCFs, with corresponding confidence intervals based on $\widehat{\textnormal{Var}}[\widehat{\mathbb{E}}[\Delta\rvcost(\evt)]]$, can be plotted similarly to MCFs. If at a given time point these intervals do not enclose zero we conclude that the difference is statistically significant to the given degree of confidence. This comparison is plotted in Figure~\ref{fig:2x2} for session and purchase counts. We see that players in Netherlands eventually play 0.5 sessions more than in Australia, on average, but this difference is not significant. In contrast, the average player in Australia makes over 0.05 purchases more, a 7-fold difference which is statistically significant. We emphasize that the test based on the MCF C.I.s is pointwise: it tells us the expected difference and the probable region of this difference for every time point, but it does not imply that if there is some time point over a long domain that the difference there must then be significant. Ideally the time $\evt$ of the test should be decided in advance.

\begin{table}
\caption{P-values of the two-sample test.}
\begin{tabu} to \linewidth {|X|X|X|}
\hline
Equality Test&Session&Purchase\\
\hline
Robust&0.56&3e-06\\
\hline
\end{tabu}
\label{table:p_values}
\end{table}

A more powerful approach compares the cohort data sets over their entire range. The two-sample MCF comparison is formulated in Cook and Lawless \cite{cook2007} as a family of test statistics as follows. For a player $i$ in cohort $k$, denote the cost $\dcost_{ki}(\evt)$ and the observable indicator $\cind_{ki}(\evt)$. Further, denote the total cost in the cohort by $\dcost_{k{\circ}}(\evt)$ and the total number of players observable by $\cind_{k{\circ}}(\evt)$.  Given a weight function $w(t)=\frac{\cind_{1{\circ}}(t)\cind_{2{\circ}}(t)}{\cind_{1{\circ}}(t)+\cind_{2{\circ}}(t)}$ and a maximum observable time $\tau=\max\{\tau_{ki}\}$, the score test statistic is 
\begin{align} \label{robcomp}
U(\tau)=\sum_{j:t_j\leq\tau}w(\evt_j)\left(\frac{\dcost_{2{\circ}}(\evt_j)}{\cind_{2{\circ}}(\evt_j)}-\frac{\dcost_{1{\circ}}(\evt_j)}{\cind_{1{\circ}}(\evt_j)}\right)\;,
\end{align}
The robust variance estimate for the test statistic is
\begin{align} \label{robcompvarest}
&\widehat{\textnormal{var}}[U(\tau)]\\
&=\sum_{k=1}^2\sum_{i=1}^{m_k}\left(
\sum_{j:t_j\leq\tau}w(\evt_j)\frac{\cind_{ki}(\evt_j)}{\cind_{k{\circ}}(\evt_j)}(\dcost_{ki}(\evt_j)-\frac{\dcost_{k{\circ}}(\evt_j)}{\cind_{k{\circ}}(\evt_j)})
\right)^2
\end{align}
The statistic $U(\tau)^2 /\textnormal{Var}[U(\tau)]$ is approximately chi-square distributed with one d.f. under the null hypothesis.

When we apply the test, we obtain the results in Table~\ref{table:p_values} which align with the pointwise test. The difference between session MCFs is not statistically significant but the difference between purchases is. This test is very useful because it allows the entire data set to be used to test for a possible difference. 

If all players have the same censoring time, the method again corresponds to a standard test. A shared censoring time $\tau_i=\tau$ simplifies the formulas (\ref{robcomp}) and (\ref{robcompvarest}) to a test that can be directly verified to equal the Welch's test of $\rvcost_1(\tau)=\rvcost_2(\tau)$, the general form of the well-known T-test.

\section{Research Analysis}

In this section we argue that the MCF estimate of the population mean has several advantages as a game analytics metric, which is suggested by its popularity in other fields. It allows the analyst to combine the retention rate type metrics and academic game metrics under one umbrella. Both benefit from the unbiasedness by varying censoring lengths. The MCF has a consistent interpretation and a lower variance compared to the retention rate. The metrics such as playtime and lifetime value can now be used in the same way as the retention rate.

\subsection{The Metric is Unbiased by Censoring}

The fact that the metric is unbiased by independent censoring \cite{cook2007} is important in real-time analytics. Censoring means that the player is not observable beyond a certain point, the current day at maximum. The possibility of using censored data is quite important in business intelligence, which often places higher demands on the time taken to act on the data than academic research. Each metric should be able to be calculated at the pace of the development iterations they are designed for, which in the app economy could be as little as months if not weeks. Note that the censoring time often varies by player; a method that handles this is able to pool the data from users arriving on different days. Pooling the data together results in more information and a higher confidence.

\begin{figure}[b]
\centering
\includegraphics[width=1.0\linewidth]{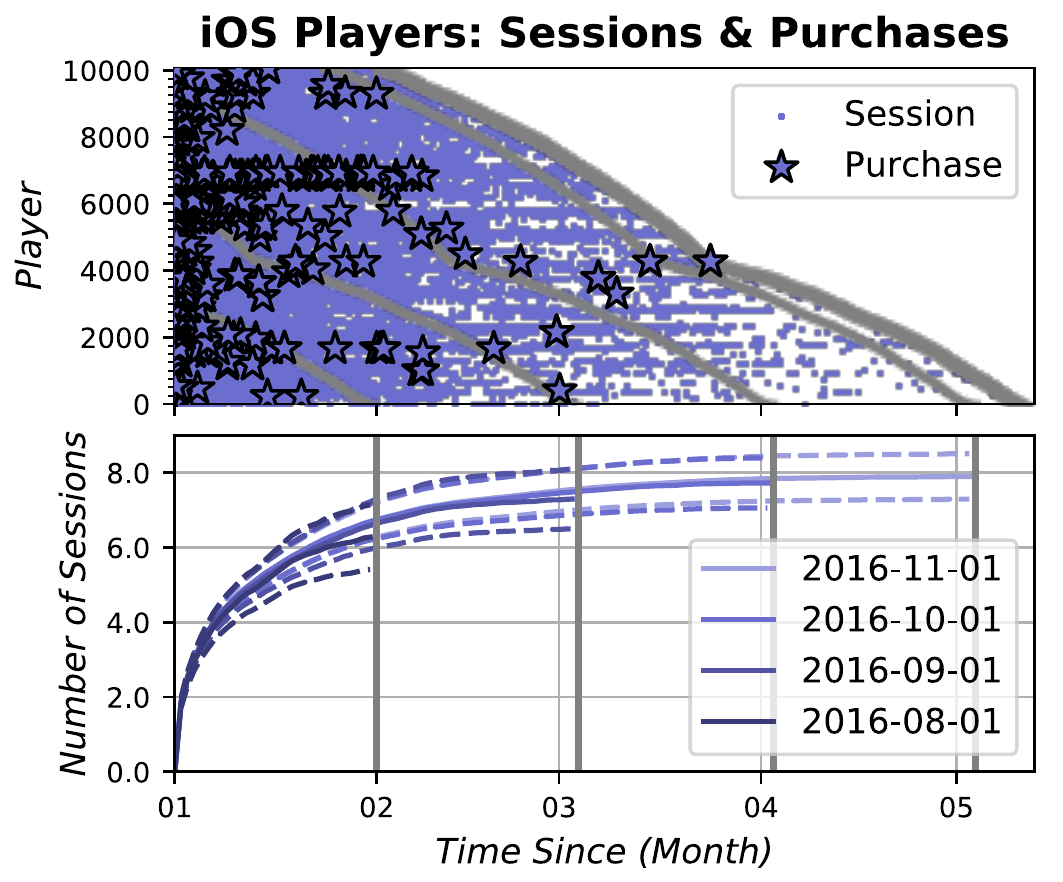}
\caption{Four MCF estimates based on different monthly censoring times. Each MCF is an unbiased estimate of the actual mean function, with the estimates becoming more accurate with a wider range as additional data arrives. }
\label{fig:censoring}
\end{figure}

The user acquisition illustrated in Figure~\ref{fig:players_population} is relatively constant, so the censoring window also varies. In Figure~\ref{fig:censoring} we have set the observation limits to the beginning of August, September, October and November of 2016, which correspond to roughly 1, 2, 3 and 4 months of data. It can be seen that an analyst who interprets the intermediate estimate correctly anticipates the actual relationship within the observable window. Time both increases the certainty within the current window and expands the window we can provide estimates for.

\subsection{The Metric Has a Consistent Interpretation}

Free-to-play game development sometimes involves trade-offs. The trade-off could be in choosing to appeal either to the most engaged or the most casual segment, resulting in a possibly a non-uniform change in retention. In Hipster Sheep for example, the Android development upgrade from the version 1.15 to 1.18 resulted in a change to retention which is displayed in Table~\ref{table:monthly_retention}. The old version had 16.52 sessions whereas the new version had 14.32 sessions per player during the first month. The long-term retention was much improved however, the new version consistently overperformed over subsequent months. The margin is quite wide, about 14-to-1 at six months. Given that one version is played more densely in the beginning and the other for longer, which is better?
\begin{table}[h!]
\caption{Recurrent sessions per player per month.}
\begin{tabu} to \linewidth {|X|X|X|X|X|X|X|X|}
\hline
Version&01&02&03&04&05&06&07\\
\hline
1.15&\textbf{16.52}&0.62&0.15&0.04&0.01&0.01&0.00\\
1.18&14.32&\textbf{1.55}&\textbf{0.78}&\textbf{0.39}&\textbf{0.23}&\textbf{0.14}&\textbf{0.04}\\
\hline
\end{tabu}
\label{table:monthly_retention}
\end{table}

\begin{figure}[b]
\centering
\includegraphics[width=1.0\linewidth]{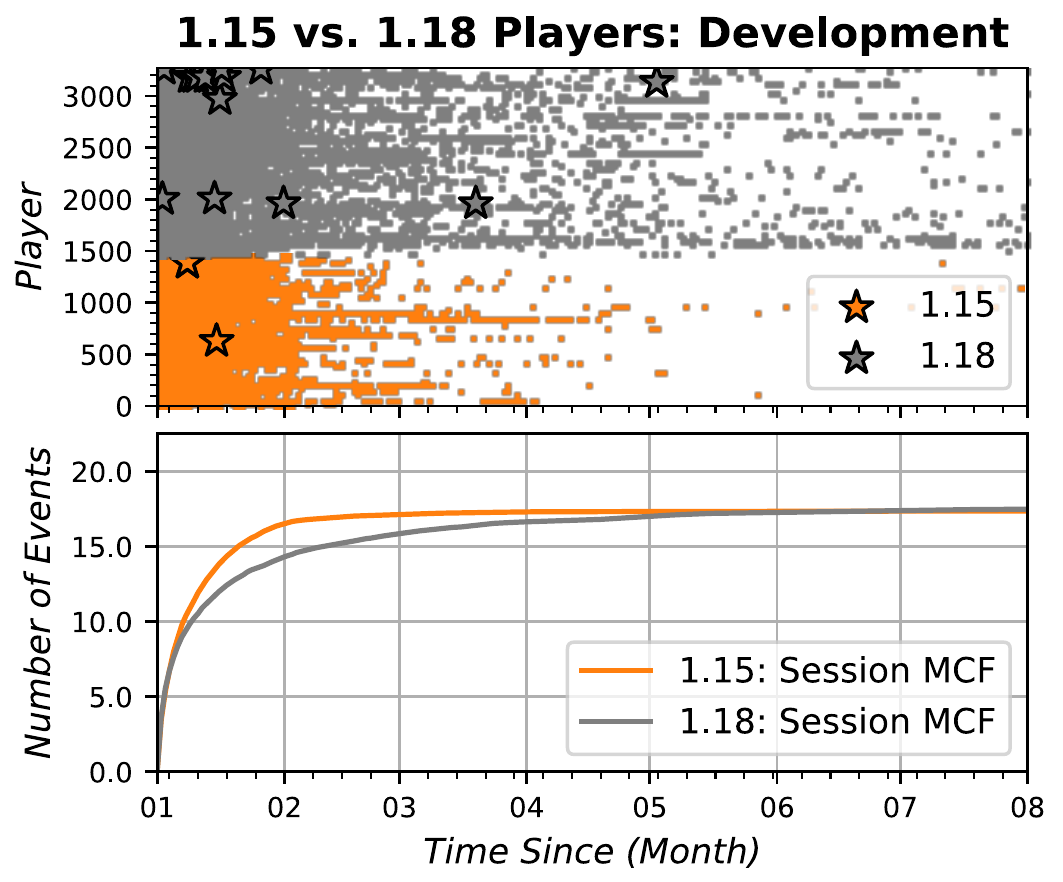}
\caption{The user test performed by comparing Android 1.15 and 1.18 versions during development is inconclusive due to shifting short- and long-term rates. The MCF asymptotes show that the trade-off actually balances eventually.}
\label{fig:118vs115}
\end{figure}

The MCF gives a simple, unambiguous measure of what the retention rate metrics imply in total. From Figure~\ref{fig:118vs115} we see that the eventual expected number sessions per player are the same between the two versions after about 6 months, leading us to conclude that the change affected the pace at which the game is played, but not the total amount of time spent playing. The cumulative sum therefore has a consistent interpretation over time, because it reveals the total effect of the individual retention rates and they can be examined as well from the derivative if desired. The total expected playtime or profit, regardless of the rate profile, can be used as a sovereign metric in a given game since the interpretation is always clear and business relevant.

\subsection{The Metric Has a Lower Sampling Variance}

One important feature of a metric is that the interpretations are robust against noise, which in this context means variation caused simply by chance. When one wants to understand user retention or monetization, an important source of noise is the random variation caused by the limited number of players. As a cumulative sum based metric, the MCF smooths the noise around the unknown base trend. This is caused by the fact that whereas random realizations of play or purchase intensity may be over or under the expected value, adding the successive realizations together causes the over- and underestimates to cancel. The metric is therefore more robust against noise.

\begin{figure}[h]
\centering
\includegraphics[width=1.0\linewidth]{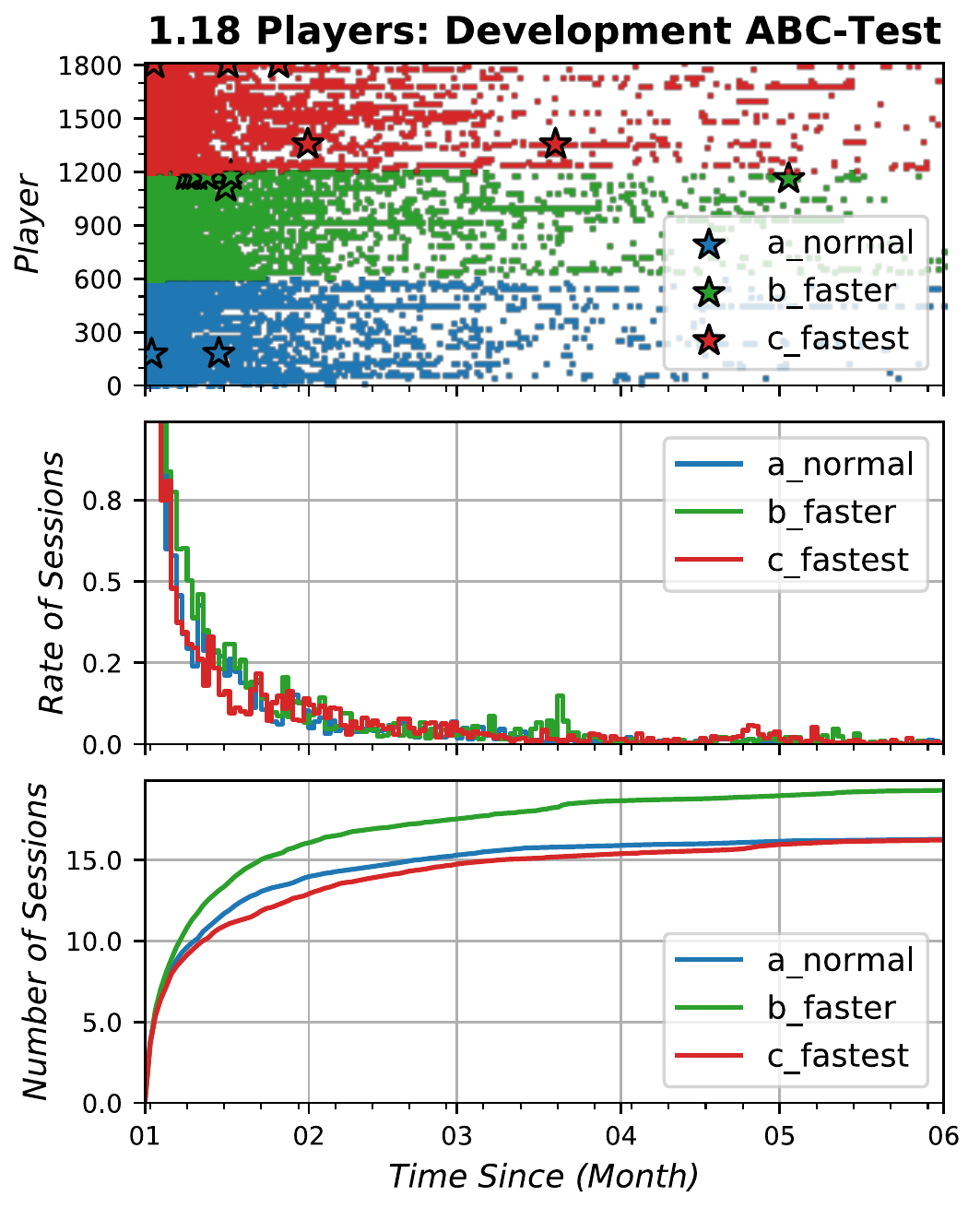}
\caption{Progression speed cohorts a\textunderscore normal, b\textunderscore faster, c\textunderscore fastest were used in Android development version 1.18 to conduct a simple ABC-test. The rate of sessions is a much noisier statistic than the cumulative number of sessions.}
\label{fig:abc_test}
\end{figure}

This is illustrated in Figure~\ref{fig:abc_test}, which plots the two different ways of understanding the results of a simple ABC-test in Android version 1.18 with 1800 players. The players were randomly assigned to three different cohorts with progression speeds normal, faster and fastest. The topmost player timelines are hard to interpret directly. If the number of returning players is plotted as the number of sessions in the middle, this retention rate metric can be seen to depend randomly on the day we pick. With only 600 players in each cohort, the metric is very sensitive to noise. However, as can be seen in the bottom of Figure~\ref{fig:abc_test}, the MCF on the other hand is smooth enough so that the rate is in fact visually easier to estimate as the slope of the MCF as it is using piecewise daily binning in the middle. This test demonstrated to us that the player retention rate may be slightly higher in the faster game version.

\begin{figure}[t]
\centering
\includegraphics[width=1.0\linewidth]{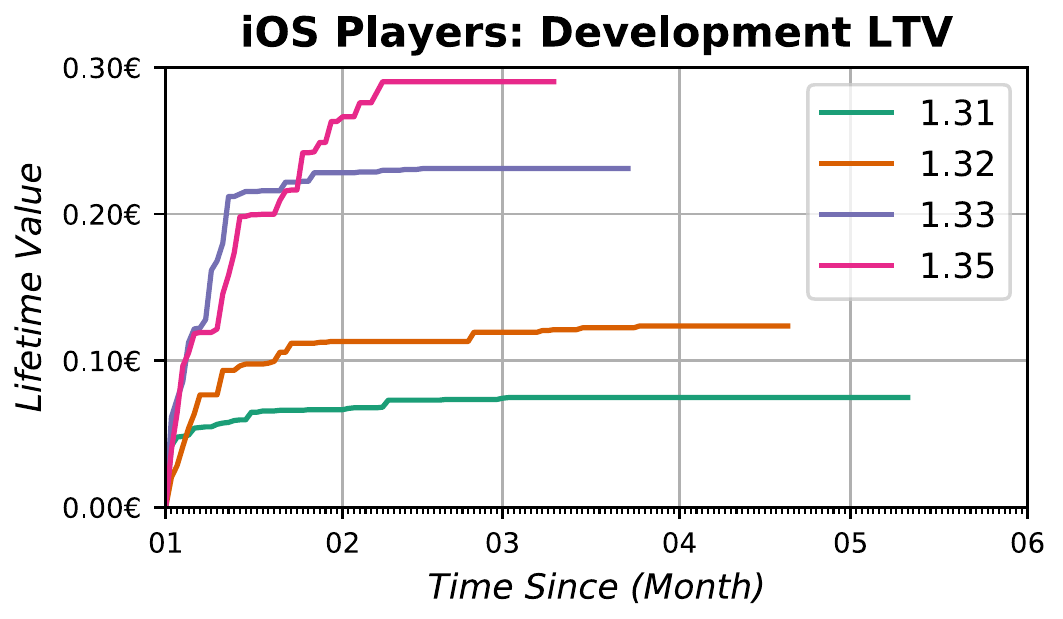}
\caption{The subsequent development effort focused on monetization, which is visible from the increasing LTV per user over small changes to the game.}
\label{fig:ios_ltv}
\end{figure}

\subsection{The Metric Can Estimate Playtime and Lifetime Value}

The final direct benefit is the versatility of the estimate. The fact that we can use the MCF with both continuous domains and ranges allows us to also estimate the expected playtime and lifetime value over time. This application in a sense generalizes the academic metrics \cite{Sifa2014,Sifa2015purchase} to censored data and allows the analyst to use them in the same way as the retention rate is currently being used. However, in this setting we are estimating the expected value and not the full distribution. This can be seen as only a small disadvantage because the expected value is often used in any case to summarize the distribution as a metric.

Because the MCF can be used more generally than the retention rate, it could have important practical implications in detecting any and all favorable changes. Many scenarios can be imagined where playtime or revenues improve but the rate metrics stay constant. For example, if the sessions become longer or purchases larger, simply counting their number over time does not detect this improvement. If we use the ordinary retention rate that counts the distinct number of daily players, even increased number of sessions or purchases within a day is not reflected in improvements to the metric.

To demonstrate that this effect is not simply hypothetical, Figure~\ref{fig:ios_ltv} for example shows how after the initial versions, during the Beta test with 10 000 iOS players, developers turned their attention to iOS monetization. They managed to improve it by small changes to the game through versions 1.31, 1.32, 1.33 and 1.35 while retention held constant. It may make sense with finished games to assume that profit is proportional to retention using a multiplier for revenue per time active, but in cases like this one needs a preferably identical tool to measure monetization at the same time as retention. The MCF is suitable for this purpose because it can measure in real-time how a continuous quantity, like euros in this case, accumulates.

\section{Conclusion}

The purpose of this study was to investigate how more efficient metrics could benefit game analytics. It was found that the Mean Cumulative Function (MCF) based metrics could provide several benefits to the game companies by providing more timely, generic and reliable information.

The MCF is a model-free estimate of the population mean, and a central concept in reliability and biostatistics. Advances in the robust analysis of recurrent events and costs make it applicable to game analytics with unknown player churn. This research found that well-founded metrics based on the MCF can have a major influence on decision making compared to current metrics by providing increased knowledge to game analysts. The study is unique in the sense that MCF has not been used in game industry and therefore it provides a novel contribution to the field of game analytics. We showed how the MCF can be used to solve important practical problems in performance measurement analogous to the way the retention rate is currently used in the industry. From a theoretical perspective, this study bridged the gap between industrial and academic game analytics by providing metrics that expanded and complemented the academic studies and sought to increase the understanding of these metrics in the game industry.

Several academic free-to-play game metrics can be generalized to censored data using the MCF and its derivative. In particular, the retention rate corresponds to the derivative when the MCF is applied to the number of distinct days played. The academic metrics of playtime and lifetime value can now be used in the same way as the retention rate, with the MCF asymptote equal to the uncensored data set. We illustrated that the MCF also has advantages over traditional metrics in censoring, interpretation and variance. Furthermore, the MCF also provides confidence intervals for quantifying uncertainty and statistical tests for comparing cohorts. We demonstrated the arguments in a single game using different real-world game development problems, but as a fully model-free estimate the metric is essentially empirical and is therefore able to generalize to any game data. 

This study naturally has some limitations. The most important potential drawback of our method is the non-robustness of the mean, which applies to the nonparametric MCF estimate presented here and the ordinary sample mean \cite{huber2009}. The sample mean is sensitive to outliers and as an estimate of the population mean it may converge slowly if at all for certain types of distributions. If one specifically desires to estimate the expected playtime or the expected profit there is no real alternative, otherwise the median could be used. Since the free-to-play model relies to an extent on a small segment of highly profitable users, we expect that this drawback could have practical consequences for any expected monetization estimate. More research is needed to assess the reliability of the mean in gaming data to know the degree to which sample based inferences can be made in practice. The MCF would also require a parametric specification, which we have not investigate in this study, to predict outside the observed data. The asymptote of the MCF, the expected lifetime value or the expected playtime for example, is an important prediction target that can be attacked by specifying a parametric form for the curve. Extensive validation would then required because the fit depends on the data set, and as an extended topic it is more suited for future work.

\section*{Acknowledgements}

We thank Tribeflame Ltd. for their continuing participation.

\ifCLASSOPTIONcaptionsoff
  \newpage
\fi

\bibliography{myBibliography}
\bibliographystyle{IEEEtran}







\end{document}